\documentclass[twocolumn,twoside,slac_two]{revtex4}
\usepackage{epsfig,graphicx}
\usepackage{fancyhdr}
\begin{document}

\newcommand {\be}{\begin{equation}}
\newcommand {\ee}{\end{equation}}
\newcommand {\ba}{\begin{eqnarray}}
\newcommand {\ea}{\end{eqnarray}}

\renewcommand{\thesection}{\arabic{section}}
\renewcommand{\thesubsection}{\arabic{section}.\arabic{subsection}}
\renewcommand{\thefigure}{\arabic{figure}}

\title{\bf Revisiting electroweak baryogenesis in context of cancelation scenario in the MSSM }

\author{Seyed Yaser Ayazi}
\affiliation{Institute for Studies in Theoretical Physics and
Mathematics (IPM), P.O. Box 19395- 5531,Tehran, Iran.}

\begin{abstract}
We study electric dipole moment bounds on CP-violating phases in
the MSSM with special emphasis on the cancelation scenario. We
find that in the range favored by electroweak baryogenesis ({\it
i.e.,} $|\mu|\simeq M_1$ or $|\mu|\simeq M_2$), $\sin
[\theta_\mu+\theta_{M_1}]$ can be as large as ${\cal O}(1)$, even
for slepton masses below 500~GeV. Such large values of the phases
promise a successful electroweak baryogenesis. We also discuss the
possibility of large CP-odd effects at a linear collider.
\end{abstract}

\maketitle

\thispagestyle{fancy}

\section{Introduction}

It is well-known that one of the requirements to explain baryon
asymmetry of the universe, is the violation of the CP symmetry. On
the other hand elementary particles can possess Electric Dipole
Moments (EDMs), only if the CP symmetry is violated. Thus,
studying EDMs of the elementary particles can play an important
role on increasing our knowledge about creation of the baryon
asymmetry of the universe. Observing CP-violation in decay of kaon
mesons has been one of the greatest discoveries of the past
century in the elementary particle physics. This phenomenon can be
explained by the phase of the CKM mixing matrix in context of the
Standard Model (SM). This phase gives rise to $d_e\sim 10^{-38}\ e
\ {\rm cm}$ \cite{e}. The prediction of CKM phase for $d_n$ ranges
from $ 10^{-31}\ e \ {\rm cm}$  to $ 10^{-33}\ e \ {\rm cm}$
 \cite{n}. No electric dipole moment for the electron or
neutron has been  so far detected but strong bounds on these
quantities have been obtained  \cite{pdg,commins,cerncourier}
\begin{equation}
|d_e|<1.4\times 10^{-27} \ e \ {\rm cm}
 \ \ \ \ \ \ |d_n|<3.0 \times 10^{-26} \ e \ {\rm cm}.\label{10-27}
 \end{equation}
Experiments have been proposed to improve these bounds by several
orders of magnitude in a few years \cite{prospect}. But the
predictions of the SM for the EDM of electron and neutron are so
small that these bounds can not be probed in the near future
experiment.

Nonzero electric dipole moment of mercury also indicates violation
of CP. The present bound on the EDM of mercury
is\cite{Romalis:2000mg}
\begin{equation}
|d_{Hg}|<2.1\times 10^{-28} \ e \ {\rm cm}.\label{10-28}
 \end{equation}
 The CP-violating phase of the CKM matrix is not large enough to
explain the baryon asymmetry of the universe
\cite{Konstandin:2005tm}.

 An additional potential source for
CP-violation in the SM arises due to instanton effects in
QCD,~{\it i.e}, the famous $\theta_{QCD}$ \cite{Cheng:1987gp}.
However upper bound on the EDM of neutron constrain the associated
dimensionless parameter $\theta_{QCD}$-term to be less than $
2\times10^{-10}$ \cite{Peccei:1998jv} which again is too small for
explaining the Baryon asymmetry of the universe
\cite{Kuzmin:1992up}. Thus, in order to explain the baryon
asymmetry, the SM should be extended to include new sources of
CP-violation.

The Minimal Supersymmetric Standard Model (MSSM) is the most
popular way of  extending the SM as it gives a solution for
stabilizing the hierarchy problem of the SM. In the general MSSM
more than forty sources of CP violation appear \cite{godbole}.
Moreover, the MSSM in its most general form, introduces new
sources of flavor violation which can give rise to lepton flavor
violating radiative decay of leptons ( $\mu \to e \gamma$, $\tau
\to e \gamma$ and $\tau \to \mu \gamma$) and a deviation of $Br(b
\to s \gamma)$ from the prediction of the SM. There are strong
bounds on the branching ratios of the lepton flavor violating
decay \cite{LFV} as well as  on the deviation of $Br(b \to s
\gamma)$ from the prediction of the SM \cite{Gomez:2006uv}.
Motivated by this observation constrained MSSM has been developed
which reduces the number  of the parameters of the SM by imposing
universality condition at high energies. The cMSSM model is a
particularly restrictive model in that there are only two new
physical phases. These phases are associated with the phase of the
A-term (the trilinear scalar terms in the soft supersymmetry
breaking potential) and the mu-term (bilinear Higgs mass term in
the superpotential). Considering the values for cMSSM parameters
which phenomenologically are favorable, one finds that the EDMs of
the electron, neutron and mercury exceed the experimental bounds
by several orders of magnitude. In the literature, to solve this
problem three solution are proposed:

\begin{itemize}

\item cMSSM phases are either zero or very small;
\item The first generation of sleptons and the first two generations of squarks are very heavy
\cite{Nath:1991dn};
\item Different contributions of the cMSSM phases to EDMs cancel each
other.

\end{itemize}

If the phases are very small, cMSSM cannot explain baryon
asymmetry of the universe. Also if second solution is realized,
the production and study cMSSM particles at ILC and LHC will be
difficult, if possible at all \cite{rubakov,stefano}. Then from
the phenomenological point of view, first two solution are not
favorable.

The third possibility has been extensively studied in the
literature \cite{heavymasses,cancelation}. Since it ostensibly
allows for large amount of CP-violating phase with relatively
small slepton and squark masses. Unfortunately it seems that
cancelation scenario works only if the phase of $\mu$ is ${\cal
O}(10^{-2})$ or less which is too small to result in detectable
CP-violating effects in colliders. This is due to the fact that
there are only two CP-violating phases which cannot simultaneously
satisfy experimental upper bounds on $d_e$, $d_n$ and $d_{Hg}$,
and also for large $\tan \beta$ regime, the contribution of
$\theta_\mu$ to the EDMs of the electron as well as the down quark
is enhanced such that it cannot be canceled by the effect of the
phase of A-term, unless the phase of $\mu$ itself is small.

Consider situation that we have relaxed some of the universality
conditions in the MSSM. This assumption leads to reproduction of
some new phases. Therefore, possibility of cancelation between
these new phases increases. Cancelation scenario for
non-universality conditions in the parameter ranges which
$|A_i|<1$~TeV and, $\tan \beta<10$ has been studied in Ref
\cite{plehn}. Also in \cite{YaserAyazi:2006zw}, we have shown that
cancelation scenario for $|A_s|, |A_d|>1$~TeV, intermediate values
of $\tan \beta$ and large values of $\mu$-phase and $A_i$ phases
are possible. This letter is review of results which is studied in
\cite{YaserAyazi:2006zw}. In section 2, we describe our model and
some non-universal conditions which from phenomenological point of
view are interesting. In section 3, we study the possibility of
cancelation in the parameter range favored by resonant electroweak
baryogenesis. In section 4, we discuss if the cancelation opens
the possibility of large enough phases  to cause sizeable
CP-violating effects in phenomena at colliders.

\section{The model}

In this paper, we consider Minimal Supersymmetric Standard Model
with the superpotential
 \ba
   W_{MSSM} &=&  Y_{u}\widehat{{u}^c}  \ \widehat{Q} \cdot \widehat{H_{u}}
        -Y_{d} \widehat{{d}^c} \ \widehat{Q} \cdot \widehat{H_{d}}
        - Y_{e}  \widehat{{e}^c} \ \widehat{L}  \cdot \widehat{H_{d}}
       \cr &-&\mu\ \widehat{H_{u}}\cdot \widehat{H_{d}} \ea

In the above formula,
 $\widehat{{u}^c}$, $\widehat{{d}^c}$ and $\widehat{{e}^c}$ are
the chiral superfields associated with the corresponding
right-handed fields. The  soft supersymmetry breaking part of
Lagrangian, at the electroweak scale, is taken to have the form

\ba \label{MSSMsoft}\L_{\rm soft}^{\rm MSSM} &=&-\ 1/2 \left( M_3
\widetilde{g} \widetilde{g} + M_2 \widetilde{W} \widetilde{W}+ M_1
\widetilde{B}\widetilde{B} +{\rm  H.c.} \right) \cr &-&(A_{u
i}Y_{u ii} \widetilde{{u_{i}}^c} \ \widetilde{Q_{i}} \cdot
H_{u}-A_{d i}Y_{d ii} \widetilde{{d_{i}}^c} \ \widetilde{Q_{i}}
\cdot H_{d}\cr &-& A_{e i}Y_{e ii} \widetilde{{e_{i}}^c} \
\widetilde{L_{i}} \cdot H_{d} + {\rm H.c.} )- \widetilde{Q
_{i}}^{\dag} \ m_{\tilde{Q} ii}^{2}\widetilde{Q_{i}} \cr &-&
\widetilde{L_{i}} ^{\dag} \ m_{\tilde{L} ii}^{2}\widetilde{L_{i}}
- \widetilde{(u_{i}^c)} ^{\dag} \ m_{\tilde{u}
ii}^{2}\widetilde{u_{i}^c} - \widetilde{(d_{i}^c)} ^{\dag} \
m_{\tilde{d} ii}^{2}\widetilde{d_{i}^c}\cr &-& \widetilde{e_{i}^c
}^{\dag} \ m_{\tilde{e} ii}^{2}\widetilde{e_{i}^c}- \
m_{H_{u}}^{2}\ H_{u}^{\dag}\ H_{u}-\ m_{H_{d}}^{2}\ H_{d}^{\dag}\
H_{d}\cr &-&(\ b \ H_{u}\cdot H_{d}+ {\rm H.c.}),\label{soft}\ea

 where the ``$i$"
indices determine the flavor. We have relaxed universality
assumption ({\it i.e.,} $m_{\tilde{\mu}}^2\ne m_{\tilde{e}}^2 \ne
m_{H_u}^2 \ne m_{H_d}^2$). As mentioned before, flavor violating
processes put strong bounds on the absolute values of flavor
violating masses and A-term. For this reason, we have taken all
matrices in (\ref{soft}) flavor diagonal. We have defined the
A-parameters factoring out the corresponding Yukawa couplings and
so in this model, A-parameters can possess different values and
phases.

Notice that $m_{H_u}^2$, $m_{H_d}^2$ and sfermion masses are real
because of hermiticity of lagrangian. The rest of parameters in
Eq.~\ref{soft} can in general be complex. By rephasing the fields
of $H_u$ and $H_d$, we can absorb any phase in $b$. Also by
rephasing gaugino fields we can make $M_2$ real. Therefore after
performing the field rephasing, MSSM is specified by these phases:
phases of $\mu$ and A-parameters as well as $M_3$ and $M_1$ (mass
of gauginos).

As is well-known, the condition for electroweak  symmetry breaking
determines the values of $\mu$ in terms of $m_{H_d}^2$,
$m_{H_u}^2$ and $\tan \beta$. In this paper, we do not make {\it a
priori} any assumption on the values of $m_{H_u}^2$ and
$m_{H_d}^2$ so we are free to assign any value to $|\mu|$. In this
regard, our model resembles the Non-Universal Higgs Mass (NUHM)
model which  has recently received attention in the literature
\cite{nuhm}. Here we show that cancelation scenario can revive
electroweak baryogenesis for the range of parameters that slepton
and squark are light.

As mentioned earlier, we relax the condition of universality at
the GUT scale, and so we can have values of $A_e$ and $A_d$ as
large as a few TeV while keeping the sfermion masses below TeV.
 For large values of A-terms, it is required to check CCB bounds
which arises due to Color and Charge Breaking (CCB) vacua. It is
shown \cite{gunion} that (For positive $m_{H_d}^2$) to guarantee
that no CCB occurs, it is sufficient to have \footnote{ The bounds
(\ref{aebound},\ref{adbound}) are based on tree-level analysis.
However, \cite{casas} shows that loop effects do not affect these
results.} \be \label{aebound} A_e^2
<3(m_{H_d}^2+m_{\tilde{e}_L}^2+m_{\tilde{e}_R}^2)\ee and \be
\label{adbound} A_d^2
<3(m_{H_d}^2+m_{\tilde{d}_L}^2+m_{\tilde{d}_R}^2).\ee

Now we relax the unification of the masses at high energies and
take $b$ (coefficient of b-term in soft SUSY lagrangian) to be of
order of $|m_{H_u}^2|$. For large $\tan\beta$, from electroweak
symmetry breaking condition in the MSSM, we find that $m_{H_d}^2$
can be positive and large. This means larger value for $m_{H_d}^2$
increases upper bounds on $A_e$ and $A_d$. Finally, since we are
assuming that off-diagonal elements of the $A$-terms are absent
(LFV), we do not need to be concerned about the region unbounded
from below \cite{casas}.

\section{EDM bounds and electroweak baryogenesis}

One of the unsolved mysteries of nature is the large asymmetry
between amount of baryon and antibaryon in the universe. There is
general consensus that the Baryon Asymmetry of the Universe (BAU)
has been created in the early universe after the inflation. A
number of different scenarios for generating baryon-antibaryon
asymmetry have been suggested. Electroweak baryogenesis is one of
the first scenarios that explains baryon asymmetry at electroweak
scale. As pointed out by Sakharov \cite{sakharov}, any theory
which uses a particle physics model to generate BAU in the early
universe, must satisfy the three following conditions:
\begin{itemize}
\item baryon number (B) violation;
\item violation of C and CP;
\item departure from thermal equilibrium.
\end{itemize}

It is shown that within the SM, baryogenesis cannot take place
because of the lack of out-of equilibrium condition. The
electroweak phase transition was regarded as a way to satisfy
out-of equilibrium condition. The SM computations of the Higgs
thermal potential show that, for first order phase transition,
mass of Higgs must be smaller than $32$ GeV \cite{cline}, while
experiments now demand a Higgs mass $m_H \geq114$ GeV. Also both
the C and CP symmetries should be violated in order for
baryogenesis scenarios to succeed. C violation has been observed
in weak interaction in context of SM. But as mentioned earlier,
CP-violation sources in the SM are too small to account for
observed BAU.

Baryogenesis at the electroweak phase transition needs new
particles coupled to the Higgs in order to obtain both strong
phase transition and to provide extra sources of CP violation. In
context of MSSM, it is shown \cite{cline} that stop
($\widetilde{t}$) can couple strongly to the Higgs of MSSM and so
it is possible to obtain departure from thermal equilibrium. In
general, for sub-TeV sparticle mass, the upper bounds from the EDM
consideration are so stringent that render electroweak
baryogenesis unsuccessful. It is the subject of this section to
show that thanks to cancelation scenario this mechanism can still
explain the BAU even for sub-TeV sparticle masses.

In order to have strong first order phase transition in the
context of supersymmetric electroweak baryogenesis, one of the top
squarks has to be lighter than the top quark. Moreover if the
lightest neutralino to be the lightest supersymmetric particle,
this in line infer that first neutralino must be lighter than the
top quark. Also in order to have successful electroweak
baryogenesis, mass of the CP-odd Higgs boson, $m_{A^0}$, should be
relatively low ($m_{A^0}\ll $~1 TeV). Another major requirement
for having successful electroweak baryogenesis is of course having
large enough CP-violating phases. However, in \cite{stefano} it is
shown that even  for values of $\sin\theta_\mu$ as low as
$10^{-2}$ successful electroweak baryogenesis can be a possibility
provided that we are at the resonance region
\cite{ramsey,resonance} (i. e., $|\mu|\simeq |M_1|$ or
$|\mu|\simeq |M_2|$).

\begin{figure}
\includegraphics[scale=0.7]{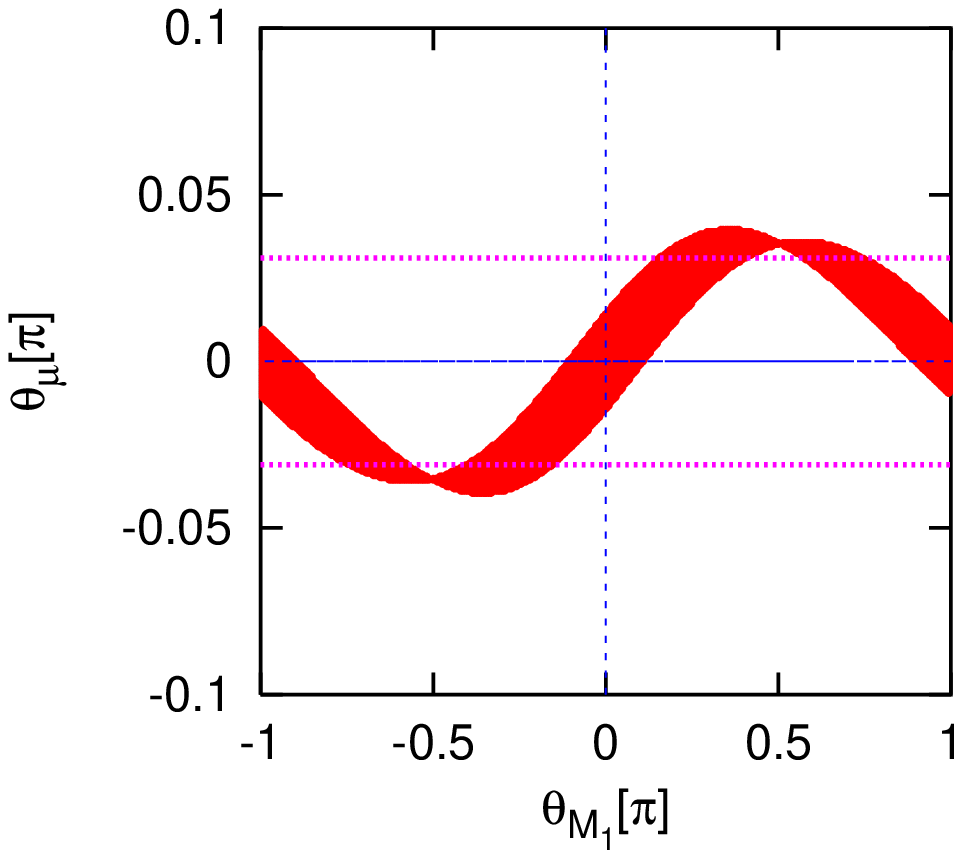} \caption{The range of phases of $\mu$ and
$M_1$ for which total cancelation among the contributions of the
phases $\mu$, $M_1$ and $A_e$ to $d_e$ is possible. We have taken
$m_{\tilde{e}_L}=392$~GeV, $m_{\tilde{e}_R}=220$~GeV,
$m_{\tilde{\nu}_L}=385$~GeV, $|A_e|=700$~GeV, $|M_1|=200$~GeV and
$M_2=415$~GeV and $\tan \beta=10$. We have set $|\mu|=200$~GeV$=
|M_1|$ which corresponds to the neutralino-driven resonance
condition of electroweak baryogenesis. The horizontal dotted lines
depict $\sin \theta_\mu=\pm 0.1$.} \label{baryogenesis200}
\end{figure}

\begin{figure}
\includegraphics[scale=0.7]{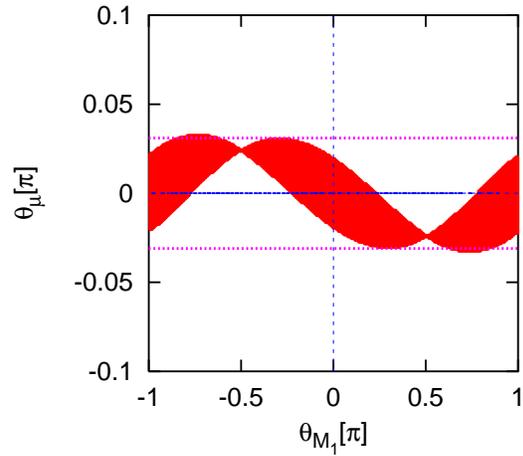} \caption{The range of phases of $\mu$ and
$M_1$ for which total cancelation among the contributions of the
phases $\mu$, $M_1$ and $ A_e$ to $d_e$ is possible. We have taken
$m_{\tilde{e}_L}=333$~GeV, $m_{\tilde{e}_R}=187$~GeV,
$m_{\tilde{\nu}_L}=324$~GeV, $|A_e|=700$~GeV, $|M_1|=167$~GeV and
$M_2=348$~GeV and $\tan \beta=10$. We have set
$|\mu|=340$~GeV$\simeq M_2$ which corresponds to the
chargino-driven resonance condition of electroweak baryogenesis.
The horizontal dotted lines depict $\sin \theta_\mu=\pm
0.1$.}\label{baryogenesis340}
\end{figure}

Note that if the masses of selectron and sneutrino are below the
TeV scale, even values of $\sin\theta_\mu$ as low as $10^{-2}$
will not be compatible with the bounds on the electric dipole
moment of electron, unless the cancelation scenario is at work.
Suppose future experiments (the LHC and ILC) confirm supersymmetry
and find out that $m_{\tilde{\chi}_1^0}<m_{\tilde{t}_R}<m_t$ and
discover a relatively light $A^0$. These conditions are
tantalizingly close to the requirement for a successful
electroweak baryogenesis. Now, suppose that the masses of
selectrons turn out to be at the scale of few hundred GeV. Does
this mean that the electroweak baryogenesis is ruled out? Figs
(\ref{baryogenesis200},\ref{baryogenesis340}) try to address this
question by studying the possibility of cancelation between
different contributions to $d_e$.

In this analysis, we consider only the one-loop effects, because
in range of the parameters which the mass of sfermions are below
the TeV scale, the dominant effects come from  one-loop. Taking
the two-loop effects into account only slightly shifts the
cancelation point.

Fig.~\ref{baryogenesis200} shows the range of phases of $\mu$ and
$M_1$ for which total cancelation among the contributions of the
phases $\mu$, $M_1$ and $A_e$ to $d_e$ is possible. To draw this
figure, we have set $\tan \beta =10$, $m_{\tilde{e}_L}=392$~GeV,
$m_{\tilde{e}_R}=218$~GeV, $m_{\tilde{\nu}_L}=385$~GeV, and
$M_2=415$~GeV. Moreover we have set  $|M_1|=|\mu|=200$ GeV which
means we are in the neutralino-driven resonant electroweak
baryogenesis regime \cite{stefano}. For this choice of parameters
the lightest neutralino is indeed lighter than the top quark. We
have set $A_e=700$~GeV which is smaller than
$[3(m_{\tilde{e}_L}^2+m_{\tilde{e}_R}^2)]^{1/2}$ thus,  as long as
$m_{H_d}^2$ is positive \cite{gunion}, no CCB will take place (see
Eq.~\ref{aebound}). Positiveness of $m_{H_d}^2$ sets a lower bound
on $b\tan\beta\simeq m_{A^0}^2$ which for our choice of parameters
is 190~GeV. Thus, for these parameters $A^0$ (the CP-odd Higgs
boson) can  still be sufficiently light. Increasing $A_e$ the
cancelation can of course become possible for larger values of
$\theta_\mu$ but the lower bound on $m_{A^0}$ will  also be
stronger and on the other hand, for heavier $m_{A^0}$ the produced
baryon asymmetry is suppressed. As shown in \cite{stefano}, the
neutralino-driven resonant baryogenesis is only marginally
compatible with the indirect searches of dark matter so this
choice of parameters in near future will be tested not only by
collider data but also by further indirect dark matter searches.

From Fig.~\ref{baryogenesis200}, we observe that for universal
gaugino masses [$\theta_{M_1}=\theta_{M_2}=0$], cancelation can
take place even for values of  $|\sin \theta_\mu|$ up to  0.06
which according to \cite{stefano} can easily yield the
baryon-antibaryon asymmetry compatible with the WMAP results. This
confirms the results of \cite{ramsey}. In the neutralino-driven
electroweak baryogenesis regime, the combination of the phases
which determines baryogenesis is $\theta_\mu+\theta_{M_1}$. Notice
that $\theta_\mu+\theta_{M_1}$ is a rephasing invariant quantity.
Fig.~\ref{baryogenesis200}  shows that, relaxing the assumption of
the universality of the gaugino masses [$\theta_{M_1}\ne
\theta_{M_2}=0$], cancelation makes $|\sin(\theta_\mu
+\theta_{M_1})|\sim 1$ compatible with the bounds on $d_e$.

Note that in our analysis, we consider only EDM of electron. In
the case of neutron and mercury there are 6 new sources of
CP-violation in comparison with the SM [phases of $\mu$, $M_1$,
$M_3$, $A_d$ , $A_u$ and $A_s$]. Thus, there exist large degrees
of freedom which makes it possible for CP-violation phases to take
large values in cancelation scenario.

 Now let us discuss fine tuning required for
such cancelation.  If the phases of $\mu$ and $M_1$ are at the
region where cancelation can take place, the generic value of
$d_e$ is already around $10^{-26}$~e cm so to reduce the value of
$d_e$ down to below the upper bound on it (see Eq.~\ref{10-27}), a
cancelation of $10\%$ will be enough which means the fine tuning
of the phases is not a problem. In Ref \cite{YaserAyazi:2006zw}
fine tuning required for the cancelation scenario in the $d_n$ and
$d_{Hg}$ have also been discussed and shown that it depends on
formulas for $d_n$, $d_{Hg}$ in terms EDMs of quarks. Also fine
tuning required for the cancelation scenario in these cases can be
stronger than EDM of electron.

In the near future, the experiments are going to become sensitive
to even smaller values of $d_n$, $d_{Hg}$ and $d_{e}$. Moreover,
there are proposals to probe EDM of deuteron down to $(1-3)\times
10^{-27}$~e~cm \cite{dD}. If one or more of these experiments
detect a nonzero electric dipole moment, it will be a strong hint
in favor of  the electroweak baryogenesis. On the other hand, if
they all report null results, we cannot still rule out the
cancelation scenario even though a new piece of information (the
bound on $d_D$) is added. However a greater degree of fine tuning
would be necessary for the cancelation.

 Fig.~\ref{baryogenesis340}
explores the possibility of cancelation scenario and having large
CP-violating phases in the chargino-driven resonant electroweak
baryogenesis regime ($|\mu|\simeq M_2$). The above discussion
holds in this case, too, with the difference that for the
chargino-driven electroweak baryogenesis the combination of phases
that are relevant for baryogenesis is $\theta_\mu+\theta_{M_2}$.
[Notice that $\theta_\mu+\theta_{M_2}$ is a rephasing invariant
quantity which in the basis we have chosen ($\theta_{M_2}=0$)
corresponds to $\theta_\mu$.] According to this figure
$\sin(\theta_\mu)$ can reach 0.1 which may be enough for a
successful baryogenesis \cite{stefano}. In case of Fig.
\ref{baryogenesis340}, since $|\mu|$ is larger, the lower bound on
the CP-odd Higgs boson will be stronger: $m_{A^0}>335$~GeV. Unlike
the case of neutralino-driven electroweak baryogenesis,
 the chargino-driven electroweak baryogenesis is not sensitive to the indirect dark matter
searches.

\section{Implication of cancelation scenario for CP-violation searches in the colliders}
The CP-violating phases can gives rise to both CP-even and CP-odd
phenomena at the LHC \cite{godbole,cplhc} and International Linear
Collider, ILC \cite{cpilc,kittelthesis,kittel}. Due to high
precision and capability of polarizing the initial beams, the ILC
will have a greater chance to observe CP-violation in the
production and decay of sparticles. In \cite{kittelthesis,kittel},
it is shown that even small values of CP-violating phases can
result in an asymmetry between $e^+e^- \to \tilde{\chi}_1^0
\tilde{\tau}_1^+ \tau^-$ and $e^+e^- \to\tilde{\chi}_1^0
\tilde{\tau}_1^- \tau^+$. Following \cite{kittelthesis}, let us
define

\begin{equation} A_{CP}\equiv {P_2-\bar{P}_2 \over 2}. \label{A-cp}\end{equation}

\begin{figure}
\includegraphics[scale=0.8]{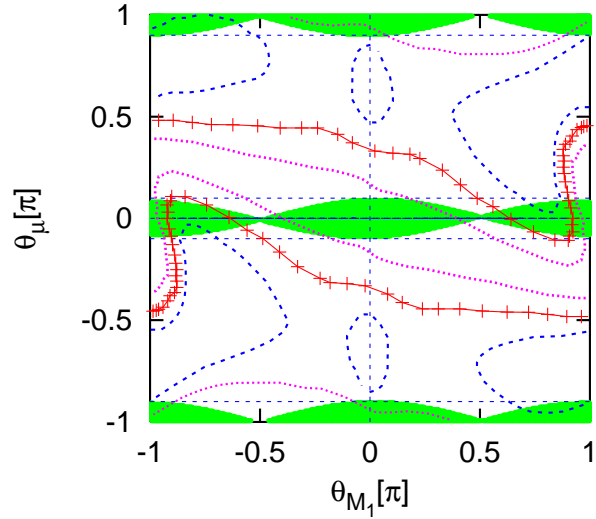} \caption{Shadowed areas show the region where
cancelation can yield vanishing $d_e$. The curves, which are
borrowed from Fig. 2.12.b of \cite{kittelthesis},
 correspond to various values of $A_{CP}$ (see the text for  the definition of $A_{CP}$):
 Dashed lines correspond to $A_{CP}=\pm 45$\%; curves marked with $+$ indicate $A_{CP}=\pm 30$\%
 and finally the thin curves correspond to $A_{CP}=\pm 15$ \%. To draw the shadowed area we have
 used the same input parameters as in Fig 2.12.b of \cite{kittelthesis}:
 $|\mu|=300$~GeV,
$m_{\tilde{e}_L}=378$~GeV, $m_{\tilde{e}_R}=211$~GeV,
$m_{\tilde{\nu}_L}=370$~GeV, $|M_1|=192$~GeV, $M_2=400$~GeV,
$|A_e|=2000$~GeV and $\tan \beta=5$. The horizontal dotted lines
correspond to $\theta_\mu=-0.9 \pi, -0.1 \pi, 0.1 \pi$ and $0.9
\pi$. } \label{figkittel}
\end{figure}

In the above definition, $P_2$ is the polarization of $\tau$ which
is produced in the subsequent processes $e^+ e^- \to
\tilde{\chi}_1^0 \tilde{\chi}_i^0$ and $\tilde{\chi}_i^0 \to
\tau^-\tilde{\tau}^+$. The polarization vector is defined as
\begin{equation}
\vec{P}\equiv \frac{{\rm Tr}[ \rho \vec{\sigma}]}{{\rm Tr}[\rho]},
\end{equation}
where $\rho$ is the spin density of $\tau$ and direction 2 is
taken to be perpendicular to the plane defined by the momenta of
the $\tau$ and the initial electron.  Curves in Fig.
\ref{figkittel}, which are borrowed  from Fig. 2.12.b of
\cite{kittelthesis}, show different contour lines corresponding to
various values of $A_{CP}$. The input data for the curves are
$\theta_{A_\tau}=0$, $A_\tau=250$~GeV and
($P_{e^-},P_{e^+})=(-0.8,0.6)$. The rest of the input parameters
are given in the caption of Fig. \ref{figkittel}. Notice that  the
input parameters satisfy the relations that we would have expected
in the mSUGRA. It is remarkable that  $A_{CP}=\pm 45\% $ can be
possible for values of $\theta_\mu$ as small as $\pm 0.1 \pi$ and
$\theta_{M_1}=\pm 1/6\pi$ or for $\theta_\mu=\pm 0.06 \pi$ and
$\theta_{M_1}=\pm \pi/2$. The shadowed areas superimposed on the
curves show the region for  which the cancelation scenario can
result in vanishing $d_e$. In order to check if in the same area
vanishing $d_n$ and $d_{Hg}$ is possible, we  calculated the
corresponding gluino and squark masses in the specific point in
the mSUGRA space chosen above and inserted them in the formulae
for $d_{Hg}$ and $d_n$. We found that for any given set of
$\theta_\mu$ and $\theta_{M_1}$ total cancelation can
simultaneously suppress the values of $d_n$ and $d_{Hg}$. The
overlap of curves with the shadowed area indicates that even for
light sfermion masses, we still have a hope to observe
CP-violating effects at ILC provided that the systematic and
statistical errors are under control.

Let us now discuss the fine tuning required for suppressing the
EDMs below the upper bounds on them. Taking $\theta_\mu$ and
$\theta_{M_1}$ in  the shadowed area and assigning a general value
between $-\pi$ and $\pi$ to $\theta_{A_e}$ we find that $d_e$
cannot exceed $10^{-26}$~e cm. This means that the fine tuning
required to suppress $d_e$ below the bound in Eq. \ref{10-27} is
not greater than 10\%. However, although simultaneous suppression
of $d_n$ and $d_{Hg}$ is possible for a wider range of phases, we
have found that the required fine-tuning in this case is greater
and is of order of 1\%.

\section{Conclusions}
In this letter, we have studied the the cancelation scenario for
EDMs in the context of general MSSM in the parameter range that is
interesting from phenomenological point of view ({\it i.e.}, low
sparticle masses). In our model, we assumed some of the conditions
which phenomenologically are interesting. We have studied the
possibility of cancelation for the region that electroweak
baryogenesis is enhanced ($|\mu| \simeq |M_1|$ and $|\mu|\simeq
|M_2|$) and found that, even for the sub-TeV slepton masses,
$|\sin \theta_\mu|\simeq 0.1$ and $|\sin \theta_{M_1}| \simeq 1$
can be compatible with the EDM bounds. The main point is that
relaxing the assumption of the universality of gaugino mass phases
$(\theta_{M_1}\ne \theta_{M_2})$ makes more effective cancelation
such that values of $\left| \sin [\theta_\mu +\theta_{M_1}]\right|
\sim 1$ become compatible with the bounds on $d_e$. This opens new
windows towards successful electroweak baryogenesis.  Notice that
in this range of parameters the fine-tuning required for
successful cancelation is not too high.

We have then focused on CP-odd quantities associated with the
decay of neutralinos at ILC and have found that thanks to
cancelation, $A_{CP}$ ( see the definition in Eq.~\ref{A-cp}) as
large as $45\%$ can be possible.

\section{Acknowledgement}
I would like to thank the organizer of IPM School and conference
on lepton and hadron where this talk was presented. I am specially
grateful to Y. Farzan for careful reading of manuscript, and for
the comments. I also appreciate M. Nikamal for teaching me about
Latex software.

\end{document}